\documentclass[twocolumn,pre,superscriptaddress]{revtex4}
\usepackage{graphics}
\usepackage{graphicx}
\usepackage{amsfonts}
\usepackage{textcomp}
\usepackage{amssymb}
\usepackage{mathrsfs}
\usepackage{amsmath}
\usepackage{color}
\usepackage{ulem}

\begin{document}

\title{Enhanced Electronic Transport in Disordered Hyperuniform Two-Dimensional Amorphous Silica}

\author{Yu Zheng\footnote{These authors contributed equally to this work.}}
\affiliation{Department of Physics, Arizona State University,
Tempe, AZ 85287}
\author{Lei Liu\footnotemark[1]}
\affiliation{Materials Science and Engineering, Arizona State
University, Tempe, AZ 85287}
\author{Hanqing Nan\footnotemark[1]}
\affiliation{Materials Science and Engineering, Arizona State
University, Tempe, AZ 85287}
\author{Zhen-Xiong Shen\footnotemark[1]}
\affiliation{Key Laboratory of Quantum Information, University of
Science and Technology of China, Hefei, Anhui 230026, P. R. China}
\affiliation{Synergetic Innovation Center of Quantum Information
and Quantum Physics, University of Science and Technology of
China, Hefei, Anhui 230026, P. R. China}
\author{Ge Zhang}
\affiliation{Department of Physics, University of Pennsylvania,
Philadelphia, PA 19104}
\author{Duyu Chen}
\affiliation{Tepper School of Business, Carnegie Mellon
University, Pittsburgh, PA 15213}
\author{Lixin He}
\affiliation{Key Laboratory of Quantum Information, University of
Science and Technology of China, Hefei, Anhui 230026, P. R. China}
\affiliation{Synergetic Innovation Center of Quantum Information
and Quantum Physics, University of Science and Technology of
China, Hefei, Anhui 230026, P. R. China}
\author{Wenxiang Xu}
\email[correspondence sent to: ]{xwxfat@gmail.com}
\affiliation{College of Mechanics and Materials, Hohai University,
Nanjing 211100, P.R. China} \affiliation{Materials Science and
Engineering, Arizona State University, Tempe, AZ 85287}
\author{Mohan Chen}
\email[correspondence sent to: ]{mohanchen@pku.edu.cn}
\affiliation{CAPT, HEDPS, College of Engineering, Peking
University 100871, P.R. China}
\author{Yang Jiao}
\email[correspondence sent to: ]{yang.jiao.2@asu.edu}
\affiliation{Materials Science and Engineering, Arizona State
University, Tempe, AZ 85287} \affiliation{Department of Physics,
Arizona State University, Tempe, AZ 85287}
\author{Houlong Zhuang}
\email[correspondence sent to: ]{hzhuang7@asu.edu}
\affiliation{Materials Science and Engineering, Arizona State
University, Tempe, AZ 85287}

\begin{abstract}
Disordered hyperuniformity (DHU) is a recently proposed new state
of matter, which has been observed in a variety of classical and
quantum many-body systems. DHU systems are characterized by
vanishing infinite-wavelength density fluctuations and are endowed
with unique novel physical properties. Here we report the first
discovery of disordered hyperuniformity in atomic-scale 2D
materials, i.e., amorphous silica composed of a single layer
of atoms, based on spectral-density analysis of high-resolution
transmission electron microscope images. Subsequent
simulations suggest that the observed DHU is closely related to
the strong topological and geometrical constraints induced by the
local chemical order in the system. Moreover, we show via
large-scale density functional theory calculations that DHU leads
to almost complete closure of the electronic band gap compared to
the crystalline counterpart, making the material effectively a
metal. This is in contrast to the conventional wisdom that
disorder generally diminishes electronic transport and is due to
the unique electron wave localization induced by the topological
defects in the DHU state.
\end{abstract}
\maketitle


Disorder hyperuniform (DHU) systems are a unique class of
disordered systems which suppress large-scale density fluctuations
like crystals and yet possess no Bragg peaks \cite{ref1, ref2}.
For a point configuration (e.g., a collection of particle centers
of a many-body system), hyperuniformity is manifested as the
vanishing structure factor in the infinite-wavelength (or
zero-wavenumber) limit, i.e., $\lim_{k\rightarrow 0}S(k) = 0$,
where $k=2\pi/\lambda$ is the wavenumber. In this case of a random
field, the hyperuniform condition is given by $\lim_{k\rightarrow
0}\hat{\psi}(k) = 0$, where $\hat{\psi}(k)$ is the spectral
density \cite{ref2}. It has been suggested that hyperuniformity
can be considered as a new state of matter \cite{ref1}, which
possesses a hidden order in between of that of a perfect crystal
and a totally disordered system (e.g. a Poisson distribution of
points).

Recently, a wide spectrum of physical and biological systems have
been identified to possess the remarkable property of
hyperuniformity, which include the density fluctuations in early
universe \cite{ref3}, disordered jammed packing of hard particles
\cite{ref4, ref5, ref6, ref7}, certain exotic classical ground
states of many-particle systems \cite{ref8, ref9, ref10, ref11,
ref12, ref13, ref14, ref15}, jammed colloidal systems \cite{ref16,
ref17, ref18, ref19}, driven non-equilibrium systems \cite{ref20,
ref21, ref22, ref23}, certain quantum ground states \cite{ref24,
ref25}, avian photoreceptor patterns \cite{ref26}, organization of
adapted immune systems \cite{ref27}, amorphous silicon
\cite{ref28, ref29}, a wide class of disordered cellular materials
\cite{ref30}, dynamic random organizating systems
\cite{hexner2017noise, hexner2017enhanced, weijs2017mixing,
lei2019nonequilibrium, lei2019random}, and even the distribution
of primes on the number axis \cite{torquato2019hidden}. In
addition, it has been shown that hyperuniform materials can be
designed to possess superior physical properties including large
isotropic photonic band gaps \cite{ref31, ref32, ref33}, optimized
transport properties \cite{ref34}, mechanical properties
\cite{ref35} as well as optimal multi-functionalities
\cite{ref36}. Designer DHU materials have also been successfully
fabricated or synthesized using different techniques \cite{ref37,
ref38}.

In this letter, we report the discovery of hyperuniformity in
amorphous 2D silica (conventionally modeled as ``continuous random
networks'' \cite{ref39}), based on the analysis of aberration
corrected transmission electron microscopy (TEM) images of the
material. To the best of our knowledge, this is the first
discovery of disordered hyperuniformity in atomic scale 2D
materials (i.e., those composed of a single layer of atoms),
which can possess unique novel electronic, magnetic and optical
properties compared to their bulk counterparts.

We show via simulations that the observed DHU in amorphous
silica is closely related to the strong topological and
geometrical constraints induced by the local chemical order in the
system. In addition, our density functional theory calculations
show that DHU significantly reduces the electronic band gap in 2D
amorphous silica, leading to almost complete closure of the band
gap compared to the crystalline counterpart. This is in contrast
to the conventional wisdom that disorder generally diminishes
electronic transport and is due to the unique electron wave
localization induced by DHU.

\begin{figure}[ht]
\includegraphics[width=0.45\textwidth,keepaspectratio]{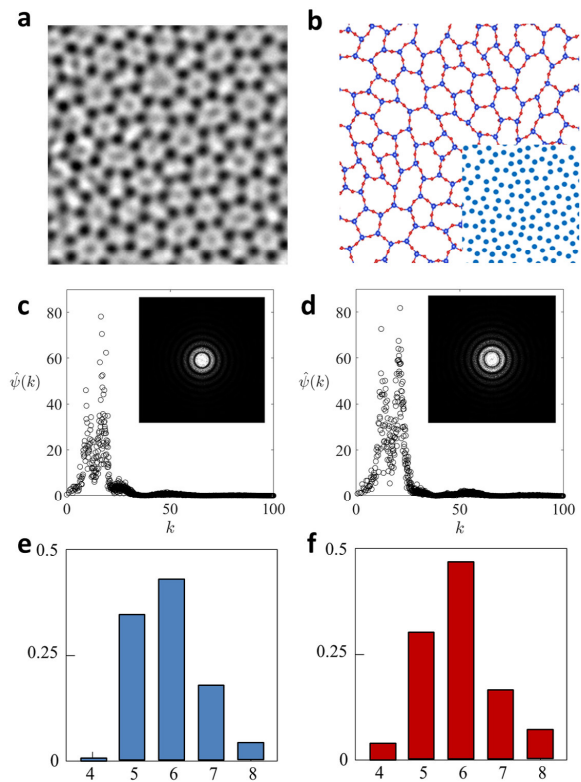}
\caption{Hyperuniformity in 2D amorphous silica. (a) TEM image of
2D amorphous silica. Reproduced from Ref. \cite{ref40} (b)
Disordered hyperuniform silica network generated via computer
simulations as described in the text. Blue and red spheres denote
Si and O atoms, respectively. Inset shows intensity map derived
from the positions of silicon atoms for spectral density analysis.
(c) and (d) respectively show the angularly averaged spectral
density $\psi({\bf k})$ associated with the TEM micrograph and the
simulated amorphous silica network. Insets show the full spectral
density. (e) and (f) respectively show the local ``ring''
statistics of the experimentally obtained and simulated 2D
amorphous silica network.} \label{fig_1}
\end{figure}

{\bf Hyperuniformity in 2D amorphous silica.} We first analyze the
high-resolution transmission electron microscopy (TEM) images of
2D amorphous silica, see Fig.\ref{fig_1}a. The materials
samples were fabricated using chemical vapor deposition (CVD) and
the procedure for obtaining the imaging data set was reported in
detail in Ref. \cite{ref40} and briefly described in the
Supporting Information (SI). As shown in Fig. 1a, the
black spots (with diffusive boundaries) represent the silicon
atoms. The micrographs are processed to retain the distribution
information of the silicon atoms by thresholding and fitting the
grayness intensity distribution associated with each silicon atom
using a Gaussian function, i.e., $G({\bf x}) = I_0 e^{-|{\bf
x}-{\bf x}_i|^2/\sigma^2}$, where $I_0$ is the maximal intensity,
${\bf x}_i$ is the center of the silicon atom and $\sigma$ is an
effective radius.

The associated spectral density $\hat{\psi}({\bf k})$ (where ${\bf
k}$ is the wave-vector) is computed following Ref.~\cite{ref41}
and shown in the inset of Fig.\ref{fig_1}c. The angularly averaged
$\hat{\psi}(k)$ (with $k = |{\bf k}|$) is shown Fig.\ref{fig_1}c.
We note that the spectral density analysis (instead of structure
factor) is employed here to efficiently utilize all of the
information on Si atom distributions contained in the TEM
micrographs and minimize the possible systematic errors induced by
converting the intensity map into center distributions. It
can be seen that $\hat{\psi}({\bf k})$ is fully isotropic and the
scattering is completely suppressed at infinite wavelength, i.e.,
$\lim_{k\rightarrow 0}\hat{\psi}(k) = 0$ with $\hat{\psi}({\bf
k})\sim k$ for small $k$ values, which indicates that the 2D
amorphous silica samples analyzed are hyperuniform (see SI for
detailed scaling analysis).


Next, we computationally generate disordered hyperuniform silica
networks. We note that unlike 3D amorphous materials such as
metallic glasses which can be simulated by numerically quenching a
high temperature liquid state, even very rapid quenching of a 2D
materials will lead to a highly crystalline materials with a small
number of local defects. Such defected crystalline materials
clearly cannot represent the experimentally obtained DHU amorphous
silicon. Together with the observation that the 2D silica
systems are HU, this motivates us to employ a structure-based
method, which is a two-step approach.

In the first step, a three-coordinated DHU network is generated
using a modified ``collective coordinate'' approach \cite{ref11},
in which a random initial configuration of points is gradually
evolved to match a prescribed targeted structure factor $S^*(k)$
while simultaneously satisfying mutual exclusion volume
constraints. The targeted $S^*(k) = 0$ for $k\le K^*$ drives the
system to a hyperuniform state and the exclusion volume
constraints ensure the final configuration can be feasibly mapped
to an amorphous silica network. In the second step, the obtained
point configuration is mapped to a three-coordinated network by
connecting a point with its three nearest neighbors. This network
is further converted to a silica network, by placing a silicon
atom centered at each point and placing an oxygen atom at the
mid-point of the two connected silicon atoms. We then perform
molecular statics simulations using the Si-O potential based on
the Tersoff parameterization \cite{ref42} as implemented in the
LAMMPS program \cite{ref43} to optimize the constructed silica
network to physically metastable states by minimizing their total
potential energy.

The spectral density of the simulated amorphous silica network
(see Fig.\ref{fig_1}b) is computed by placing a Gaussian kernel
function at the center of each silicon atom and is shown in
Fig.\ref{fig_1}d. It can be clearly seen that $\hat\psi(k)$ of the
simulated network agrees very well with the experimental data for
all wave numbers including the zero-$k$ limit, i.e.,
$\lim_{k\rightarrow 0}\hat\psi(k) = 0$ with $\hat\psi(k) \sim k$
for small $k$ values (see SI for detailed scaling analysis and
comparison to experimental results).

We note that in the second step of the simulation (i.e., the
potential energy minimization), hyperuniformity (i.e., the small
$k$ values of $S(k)$) is not constrained anymore. The evolution of
the system is dominated by the interactions of atoms and
constrained by the topology, and the positions of the atoms have
been significantly perturbed compared to the final configuration
obtained in the first step. Yet the resulting system is still
hyperuniform. This result suggests the strong geometrical and
topological constraints, i.e., the bond length and angle
associated with the Si-O bonds as well as 3-coordinated
configurations, induced by the local chemical order could lead to
the observed hyperuniformity in the system.

We also obtain and compare the number of $n$-fold rings (where
$n=3, 4, \ldots$) formed by silicon atoms in both the experimental
and simulated networks, see Fig.\ref{fig_1}(e) and \ref{fig_1}(f).
The ring statistics for the two systems again agree very well with
one another. In addition, the comparison of local structural
statistics of the experimental and numerical systems, including
the pair correlation function and nearest neighbor distribution of
the Si atoms also show excellent agreement (see SI for details).
These results indicate our numerical network model can provide a
statistically accurate structural representation of the 2D
amorphous silica system by capturing key correlations on both
large and small length scales as well as the local topological
order. Therefore, we expect that the physical properties computed
based on the numerical network model should also be representative
of those of the experimental system.




What is the origin of DHU in 2D amorphous silica? We note
that as a first approximation, the 2D silica glass can be
considered as obtained from a 2D crystalline silica network by
continuously introducing the Stone-Wales (SW) defects
\cite{stone1986theoretical}, which change the topology of the
network. However, the SW defects do not affect the number of
particles within a large observation window, since the SW
transformation is localized and only affects a pair of atoms on
the single-bond scale. Nonetheless, SW transformations of atom
pairs on the boundary of the observation window might lead to a
bounded fluctuation of particle numbers, which are scaled with the
surface area of the window. Therefore, the SW defects should
preserve hyperuniformity in the system. We provide numerical evidence for this speculation in the SI. Although the actual 2D
amorphous silica possesses a structure that deviates from the
ideal SW transferred crystalline network, the above argument could
provide a possible explanation of the observed DHU in the system.

{\bf Disordered hyperuniformity significantly reduces electronic
band gap.} We use the simulated DHU SiO$_2$ networks to calculate
its density of states (DOS) at the DFT-PBE
\cite{PhysRevLett.77.3865} level of theory. Specifically, the DHU
structure consists of three sublayers (1800 atoms; 600 Si atoms
and 1200 O atoms). For comparison, we also create a supercell of
2D crystalline SiO$_2$ with the same number of Si and O atoms.
Several models of 2D crystalline SiO$_2$ have been studied using
DFT calculations in the literature \cite{ref44}. Here we refer 2D
crystalline SiO$_2$ to the hexagonal bilayer crystalline network
observed in experiment \cite{ref45}.

We use numerical atomic orbitals \cite{chen2010systematically} as
implemented in the ABACUS package \cite{li2016large} for
calculating the electronic structure. The simulation methods and
parameters can be found in Ref.~\cite{ref46}. For comparison, we
also compute the DOS for 2D hexagonal crystalline SiO$_2$. The
energy difference between the 2D DHU and hexagonal crystalline
SiO$_2$ calculated from the Tersoff potential and DFT are both
positive and comparable (0.074 and 0.134 eV/atom, respectively).
The positive energy differences indicate that the crystalline
structure is more energetically stable than the DHU structure.
Nevertheless, the TEM image (see Fig.\ref{fig_1}(a)) shows that
the experimental atomic structure of 2D amorphous SiO$_2$ is
drastically different from the crystalline model. By contrast, the
high similarity between the DHU model and experimentally observed
atomic structure reveals the metastable nature of DHU systems.

\begin{figure}[ht]
\includegraphics[width=8cm]{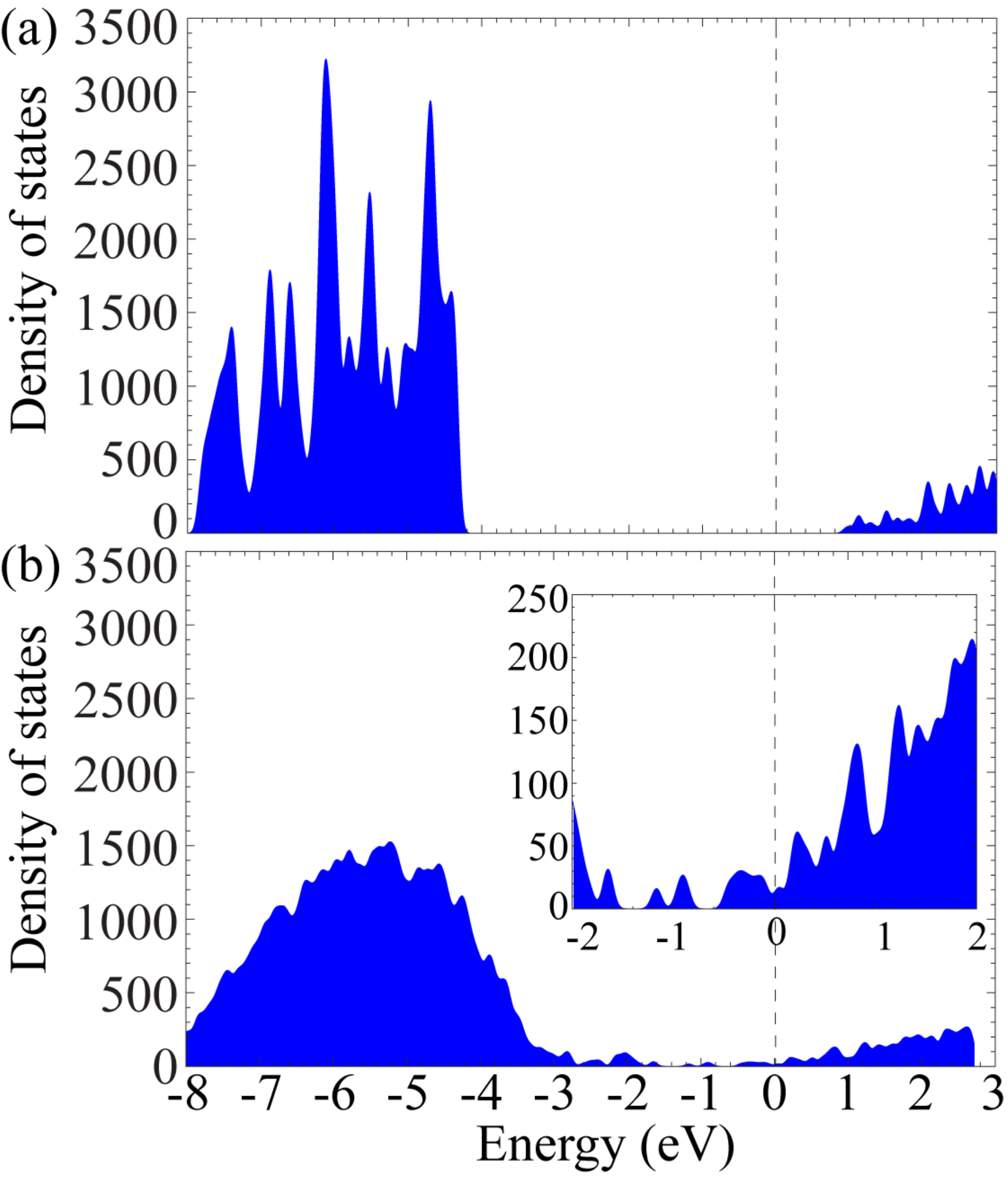}
\caption{Density of states of the supercells of 2D (a) crystalline
and (b) hyperuniform SiO$_2$ calculated using density functional
theory. The inset shows a zoom-in view of the DOS near the Fermi
level.} \label{fig:dos}
\end{figure}

Figure~\ref{fig:dos} (a) shows that the 2D crystalline SiO$_2$ is
essentially an insulator with a predicted band gap of 5.31 eV,
consistent with the previously reported band gap of 5.48 eV
calculated for a unit cell at the same level of theory
\cite{ref47}. By contrast, we observe from Fig.~\ref{fig:dos}(b)
that a small but finite number of states occupy the Fermi level of
the DHU structure, showing {\it metallic} behavior of the
electrons with a typical band gap of $\sim 50$ meV. This is
comparable to the thermal fluctuations at room temperature $\sim
25$ meV. In other words, the disordered hyperuniformity
fundamentally changes the electrical transport behavior of 2D
SiO$_2$, from an effective insulator at room temperature (as in
the crystalline form) to an effective metal (as in the DHU form).

From Fig.~\ref{fig:dos}(b), we estimate the density $\rho$ of
the electrons that contribute to the electrical conductivity of
the DHU structure at room temperature. By integrating the number
of states in the energies ranging from 25 meV (corresponding to
the thermal energy) to the Fermi level in Fig. 2(b), we determine
$\rho$ as 2.33 $\times$ 10$^{12}$ cm$^{-2}$. This magnitude belongs to
the category of ``high doping" (e.g., 6.0 $\times$ $10^{11}$ and 9.2 $\times$ $10^{12}$ cm$^{-2}$) applied to common 2D semiconductors
such as WS$_2$ and MoS$_2$ \cite{yang2014chloride}. To put it another
way, considering the electron density alone, the conductivity
associated with the DHU structure could be comparable to those of
the 2D materials.

To better understand this metallic behavior of DHU 2D SiO$_2$, we
compute the charge densities within an energy window of 0.5 eV
below the highest occupied molecular orbital (HOMO) level for the
crystalline structure and below the Fermi level for the DHU
structure. For the former structure,
Fig.\ref{fig:chargedensity}(a) (a complete version of
Fig.\ref{fig:chargedensity} can be found in the supplemental
material) shows that the electrons are distributed around Si and O
atoms in the entire structure, i.e., fully occupying the valence
bands. The amount of these electrons is significant, as can be
seen from the large DOS below the HOMO level. But these electrons
cannot be thermally excited at room temperature to the conduction
bands due to the large band gap, leading to zero electrical
conductivity for pure 2D crystalline SiO$_2$.

\begin{figure}[ht]
\includegraphics[width=8cm]{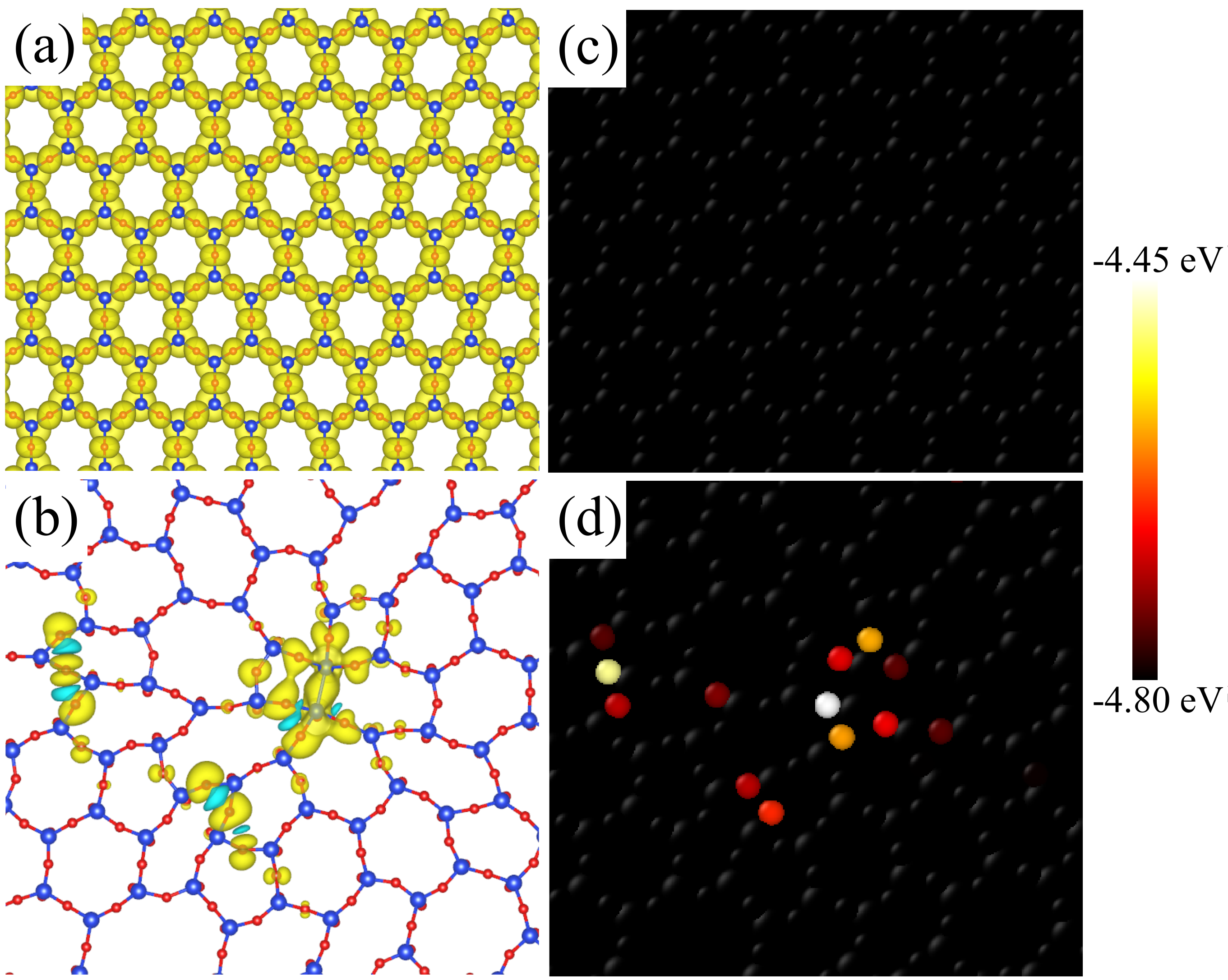}
\caption{Electron densities at the HOMO level of (a) the
crystalline structure and (b) at the Fermi level of the DHU
structure. The isosurface values is 0.0023
$e$/Bohr$^3$. The corresponding distributions of potential
energies are shown in (c) and (d), respectively. The color bar
shows a range of the distributions.} \label{fig:chargedensity}
\end{figure}

On the other hand, for the DHU structure, the number of valence
electrons in the same energy window is much less, resulting in a
low carrier density but nevertheless, a non-zero conductivity. A
closer look at the slightly wider energy window associated with
lowest density of states (e.g., -2$\sim$0 eV) reveals that the
distribution of states still forms an almost continuous spectrum
of peaks, see the inset in Fig.~\ref{fig:dos}(b).
Figure~\ref{fig:chargedensity}(b) reveals that the valence
electrons contributing to the conductivity originate from a small
portion of the Si and O atoms in the DHU system.

{\bf Topological defects in DHU SiO$_2$ lead to electron
localization with high energies.} To understand why the electrons
are localized around these atoms, we show in
Fig.~\ref{fig:chargedensity}(c) and ~\ref{fig:chargedensity}(d)
the distributions of the computed potential energies using the
Tersoff potential for the crystalline and DHU systems,
respectively. Consistent with Fig.~\ref{fig:chargedensity}(a),
Fig.~\ref{fig:chargedensity}(c) shows that the potential energies
are homogeneously distributed in the crystalline system. On the
contrary, for the DHU system, we observe from
Fig.~\ref{fig:chargedensity}(d) that the potential energies
possess a highly heterogeneous distribution, with significantly
higher energies localized in regions where the atomic arrangements
significantly deviates from the 6-fold hexagonal configuration.
These are topological defects which are necessary to achieve DHU
in amorphous 2D silica. Importantly, the high-energy localizations
perfectly coincide with the electron localizations (see
Figure~\ref{fig:chargedensity}(b)), indicating that the local
potential energies due to the topological defects induced by DHU
are sufficiently high to activate the electrons to the energy
levels near the Fermi energy.

Electron localization in a disordered system is generally
associated with a number of exotic physical phenomena \cite{ref50}
such as metal-to-insulator transitions suggested by Anderson
\cite{ref51}. The electron localization in the DHU silica system
appears to be phenomenologically different from the Anderson
localization, as the DHU localization gives rise the opposite
(i.e., insulator-to-metal) transition. A mathematical model based
on model Hamiltonian that considers the disordered potential needs
to be devised in order to investigate the electrical transport
property of the localized electrons. Furthermore, whether the
insulator-to-metal transition also occurs in other 2D DHU
semiconductors/insulators remains unknown and is certainly worth
exploring.

In summary, we have discovered, for the first time, disordered
hyperuniformity in 2D amorphous materials, and showed that DHU
fundamentally changes the electronic transport behaviors in the
material, making 2D DHU silica metallic. This interesting
prediction clearly awaits for experimental confirmation. We also
showed that the metallic behavior (i.e., with a virtually
continuous spectrum of DOS without significant band gaps in DHU
silica) is resulted from the localization of high-energy states
due to the topological defects induced by DHU. Since the observed
DHU in 2D silica is associated with the local topological and
geometrical constraints common in many 2D materials, we would
expect to observe DHU and thus, the metallic behavior, in the
amorphous states of other 2D materials, should such states be
metastable at least. With the increasing interest in 2D amorphous
materials, we expect our methods of building realistic structural
model of amorphous 2D material systems along with large-scale DFT
calculations to be applicable to a wide range of other 2D
materials such as graphene \cite{ref48} and molybdenum disulfide
\cite{ref49} in the amorphous form.

\bigskip
\noindent{\bf Acknowledgments} We thank P. Y. Huang for helpful
discussion and sharing TEM images of amorphous silica. Y. Z. and
H. N. thank Arizona State University (ASU) for the University
Graduate Fellowship. L. L. and H.Z. thank the start-up funds from
ASU. The DFT calculations have been done on the USTC HPC
facilities.


\end{document}


\title{Supplemental Material: Enhanced Electronic Transport in Disordered Hyperuniform Two-Dimensional Amorphous Silica}

\author{Yu Zheng\footnote{These authors contributed equally to this work.}}
\affiliation{Department of Physics, Arizona State University,
Tempe, AZ 85287}
\author{Lei Liu\footnotemark[1]}
\affiliation{Materials Science and Engineering, Arizona State
University, Tempe, AZ 85287}
\author{Hanqing Nan\footnotemark[1]}
\affiliation{Materials Science and Engineering, Arizona State
University, Tempe, AZ 85287}
\author{Zhen-Xiong Shen\footnotemark[1]}
\affiliation{Key Laboratory of Quantum Information, University of
Science and Technology of China, Hefei, Anhui 230026, P. R. China}
\affiliation{Synergetic Innovation Center of Quantum Information
and Quantum Physics, University of Science and Technology of
China, Hefei, Anhui 230026, P. R. China}
\author{Ge Zhang}
\affiliation{Department of Physics, University of Pennsylvania,
Philadelphia, PA 19104}
\author{Duyu Chen}
\affiliation{Tepper School of Business, Carnegie Mellon
University, Pittsburgh, PA 15213}
\author{Lixin He}
\affiliation{Key Laboratory of Quantum Information, University of
Science and Technology of China, Hefei, Anhui 230026, P. R. China}
\affiliation{Synergetic Innovation Center of Quantum Information
and Quantum Physics, University of Science and Technology of
China, Hefei, Anhui 230026, P. R. China}
\author{Wenxiang Xu}
\email[correspondence sent to: ]{xwxfat@gmail.com}
\affiliation{College of Mechanics and Materials, Hohai University,
Nanjing 211100, P.R. China} \affiliation{Materials Science and
Engineering, Arizona State University, Tempe, AZ 85287}
\author{Mohan Chen}
\email[correspondence sent to: ]{mohanchen@pku.edu.cn}
\affiliation{CAPT, HEDPS, College of Engineering, Peking
University 100871, P.R. China}
\author{Yang Jiao}
\email[correspondence sent to: ]{yang.jiao.2@asu.edu}
\affiliation{Materials Science and Engineering, Arizona State
University, Tempe, AZ 85287} \affiliation{Department of Physics,
Arizona State University, Tempe, AZ 85287}
\author{Houlong Zhuang}
\email[correspondence sent to: ]{hzhuang7@asu.edu}
\affiliation{Materials Science and Engineering, Arizona State
University, Tempe, AZ 85287}

\maketitle

\section{Experimental Procedure for Fabricating 2D Amorphous Silica Samples}
The 2D amorphous silica samples analyzed in paper were fabricated using a low-pressure chemical vapor deposition (CVD) technique which has been described in detail in Refs. [46] and [53] of the paper. In particular, the amorphous silica supported by graphene was grown on polycrystalline Cu foils attached to a quartz substrate using a low-pressure CVD technique with hexane being the precursor. The growth occurred in a quartz tube following four steps.  First, the pressure in the tube was pumped down close to zero (10$^{-2}$ mbar). Second, a formation gas (Ar/(5\%)H$_2$) with a pressure of about 5 mbar was introduced to the tube. Meanwhile, the Cu foils were heated to a high temperature (950 $^\circ$C); Third, the hexane gas with a pressure of 0.5 mbar was also introduced to the tube for 1 minute. Finally, the temperature in the tube was lowered to room temperature while the formation gas kept flowing. 
We show via simulations that the observed DHU in amorphous silica is closely related to the strong topological and geometrical constraints induced by the local chemical order in the
system. In addition, our density functional theory calculations show that DHU significantly reduces the electronic band gap in 2D amorphous silica, leading to almost complete closure of the band
gap compared to the crystalline counterpart. This is in contrast to the conventional wisdom that disorder generally diminishes electronic transport and is due to the unique electron wave localization induced by DHU.

\begin{figure*}[ht]
\includegraphics[width=0.75\textwidth,keepaspectratio]{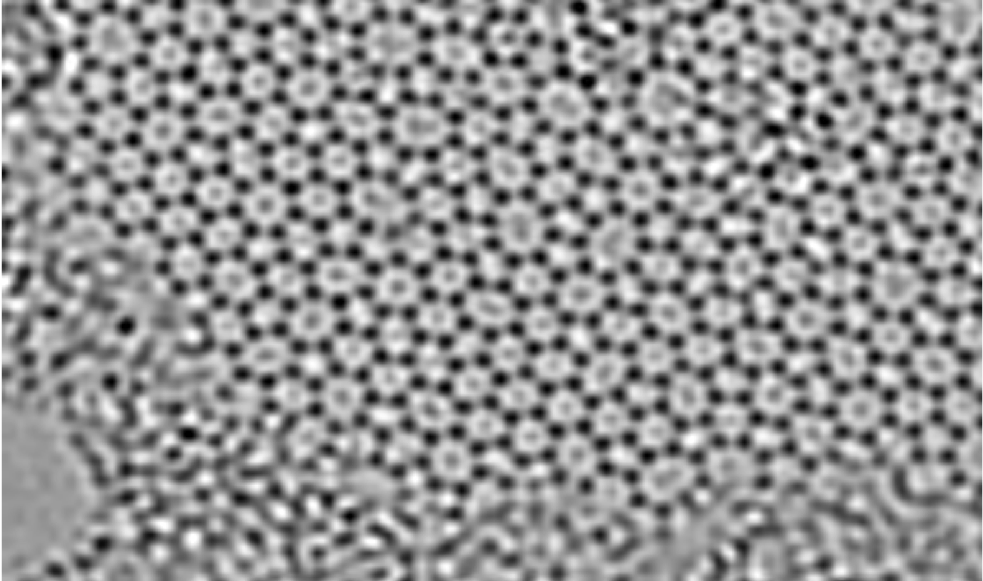}
\caption{TEM image of a typical 2D amorphous silica sample.} \label{fig_1}
\end{figure*}

It was proposed in Ref. [53] that, air leaked to the tube caused the oxidized Cu surfaces, which further reacted with the quartz substrate, leading to the formation of the amorphous silica. The material samples were further characterized and imaged using transmission electron microscope (TEM). We analyzed the TEM images of the amorphous silica samples obtained from these experimental conditions (see Fig. \ref{fig_1}). We found that all of the samples exhibit the DHU feature. Our results suggest that DHU might be an intrinsic structural feature for all amorphous 2D materials, due to the strong topological and geometrical constraints induced by local chemical orders in these systems. 

\section{Spectral Density Analysis and Hyperuniformity}
In the main paper, we provided the spectral density analysis of the experimental 2D amorphous silica and the numerical model, which show excellent agreement with one another. Here, we perform a scaling analysis of the small-$k$ behavior of the spectral density $\psi$($k$)  and analysis of hyperuniformity in the infinite-system limit (zero-wavenumber limit). 

\begin{figure*}[ht]
\includegraphics[width=0.75\textwidth,keepaspectratio]{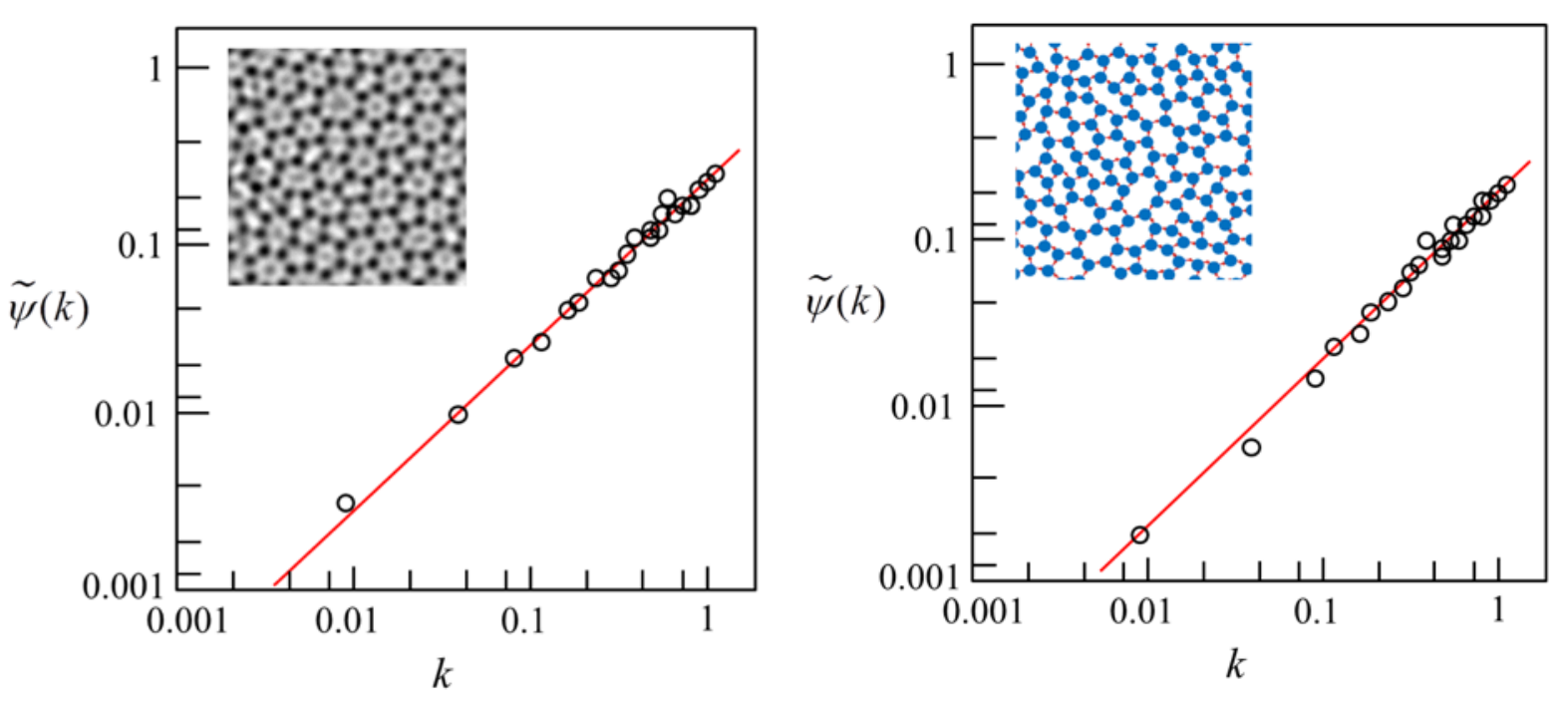}
\caption{Spectral density $\psi$($k$) at small wavenumbers in log-log plot for the experimental system (left panel) and numerical model (right panel), which indicates that spectral density scales linearly with $k$ for both systems.} \label{fig_2}
\end{figure*}

Fig. \ref{fig_2} shows the spectral density $\psi$($k$) at small wavenumbers in log-log lot for the experimental system (left panel) and numerical model (right panel). The data was fitted using linear regression, and the slope for the experimental system and numerical model is 1.021 and 1.079, respectively. These results indicate that the spectral density scales linearly with $k$ for both systems, i.e., $\psi$($k$) $\propto$ $k$. Similar scaling behaviors have also been observed in disordered jammed packing of hard particles and photoreceptor distributions in avian retina \cite{jiao2014avian,dreyfus2015diagnosing}.

\begin{figure}[ht]
\includegraphics[width=0.45\textwidth,keepaspectratio]{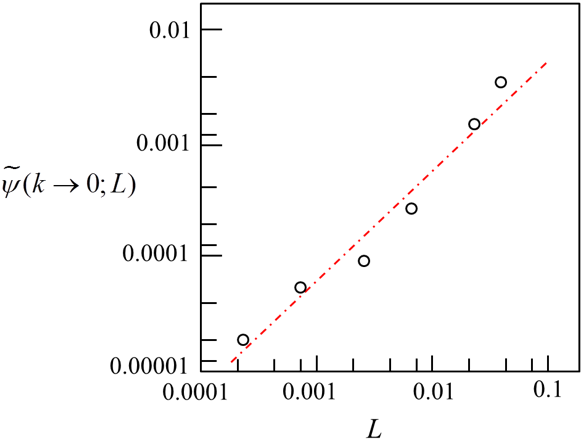}
\caption{Extrapolation of the zero-wavenumber limit of the spectral density   in the infinite-system limit (i.e., the thermodynamic limit) based on the numerical results. This scaling analysis indicates the system remains hyperuniform in the thermodynamic limit.} \label{fig_3}
\end{figure}

To ascertain hyperuniformity of the system in the thermodynamic limit, we carry out a scaling analysis of zero-wavenumber limit of the spectral density $\psi$($k \rightarrow 0$) estimated from finite systems. In particular, for a given finite system size $L$, we first estimate $\psi$($k \rightarrow 0$; $L$) by employing a 3{$^\mathrm{rd}$} order polynomial fitting, i.e., $\psi$($k$; $L$) $\approx a_0(L)+a_1(L)k+a_2(L)k^2+a_3(L)k^3$  for small $k$ values, where the coefficient   $a_0(L)$ provides an estimate of $\psi$($k \rightarrow 0$; $L$). Fig. \ref{fig_3} shows the estimated $\psi$($k \rightarrow 0$; $L$) as a function of system size 1/$L$, obtained from the numerical system. The analysis shows that $\psi$($k \rightarrow 0$; $L$) is clearly a monotonic decreasing function of $L$ and  $\psi$($k \rightarrow 0$; $L \rightarrow \infty$) $\rightarrow$ 0, which indicates the system is hyperuniform in the thermodynamic limit. We note that due to the limited field of view of the TEM, we were not able to carry out a similar scaling analysis based on the experimental data. However, since the numerical model is structurally representative of the actual system on different length scales and topologically, we expect the actual 2D amorphous silica is also hyperuniform in the thermodynamic limit. Finally, we note that increasing system size does not affect the linear scaling behavior for small wavenumbers.

\section{Comparison of Structural Statistics of Experimental System and Numerical Models}
\begin{figure*}[ht]
\includegraphics[width=0.75\textwidth,keepaspectratio]{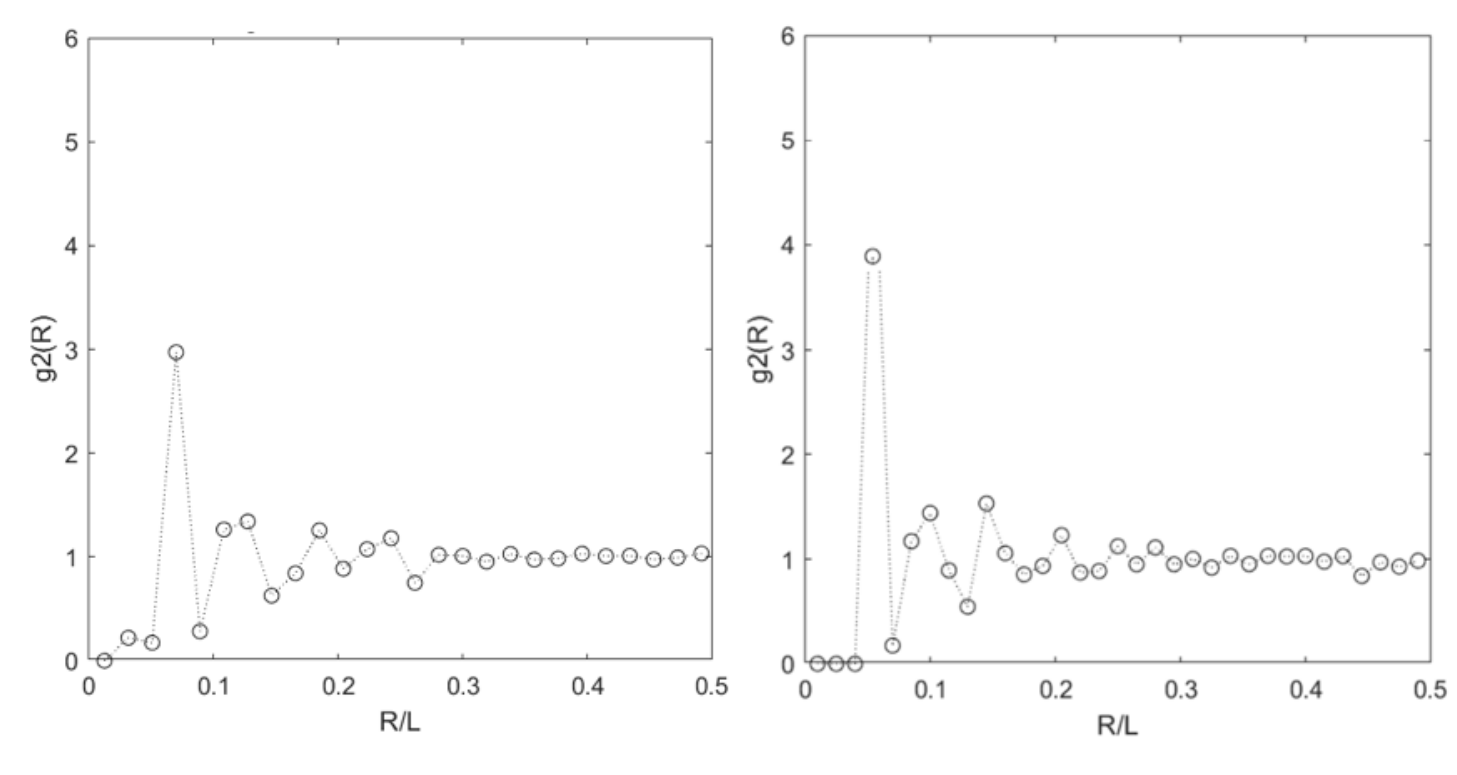}
\caption{Pair-correlation function $g_2$($r$) associated with the Si atoms of the experimental system (left) and numerical network model (right).} \label{fig_5}
\end{figure*}

\begin{figure*}[ht]
\includegraphics[width=0.75\textwidth,keepaspectratio]{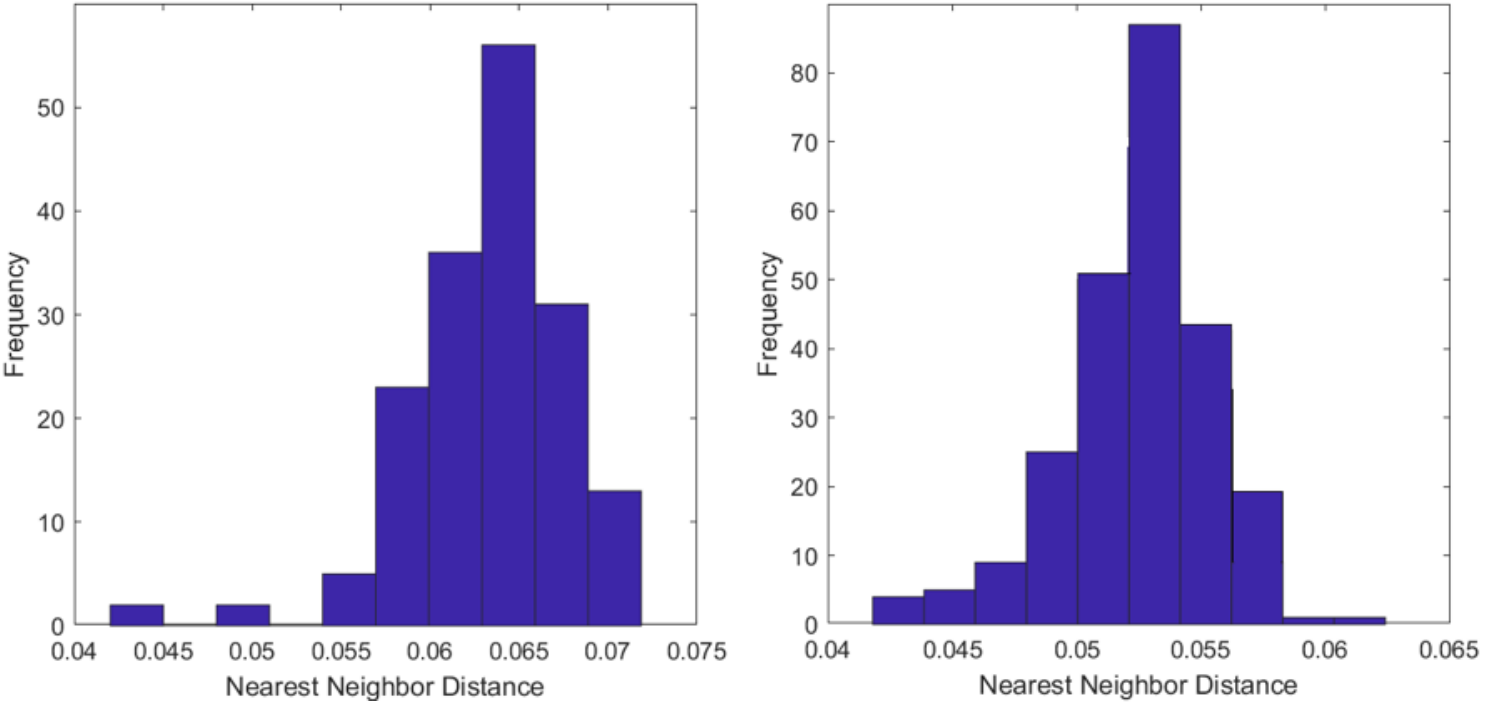}
\caption{Nearest-neighbor distribution P($r$) associated with the Si atoms of the experimental system (left) and numerical network model (right).} \label{fig_6}
\end{figure*}

In the main paper, we show that spectral density and ring statistics of the experimental system and numerical model agree very well with one another, indicating the model is structurally representative of the 2D amorphous silica on large length scales and topologically. Here, we compute and compare local structural statistics of the experimental and numerical systems, including the pair-correlation function $g_2$($r$) and the nearest-neighbor distribution P($r$) associated with the Si atoms, which are shown in Fig. \ref{fig_5} and Fig. \ref{fig_6}, respectively. 

\begin{figure*}[ht]
\includegraphics[width=0.75\textwidth,keepaspectratio]{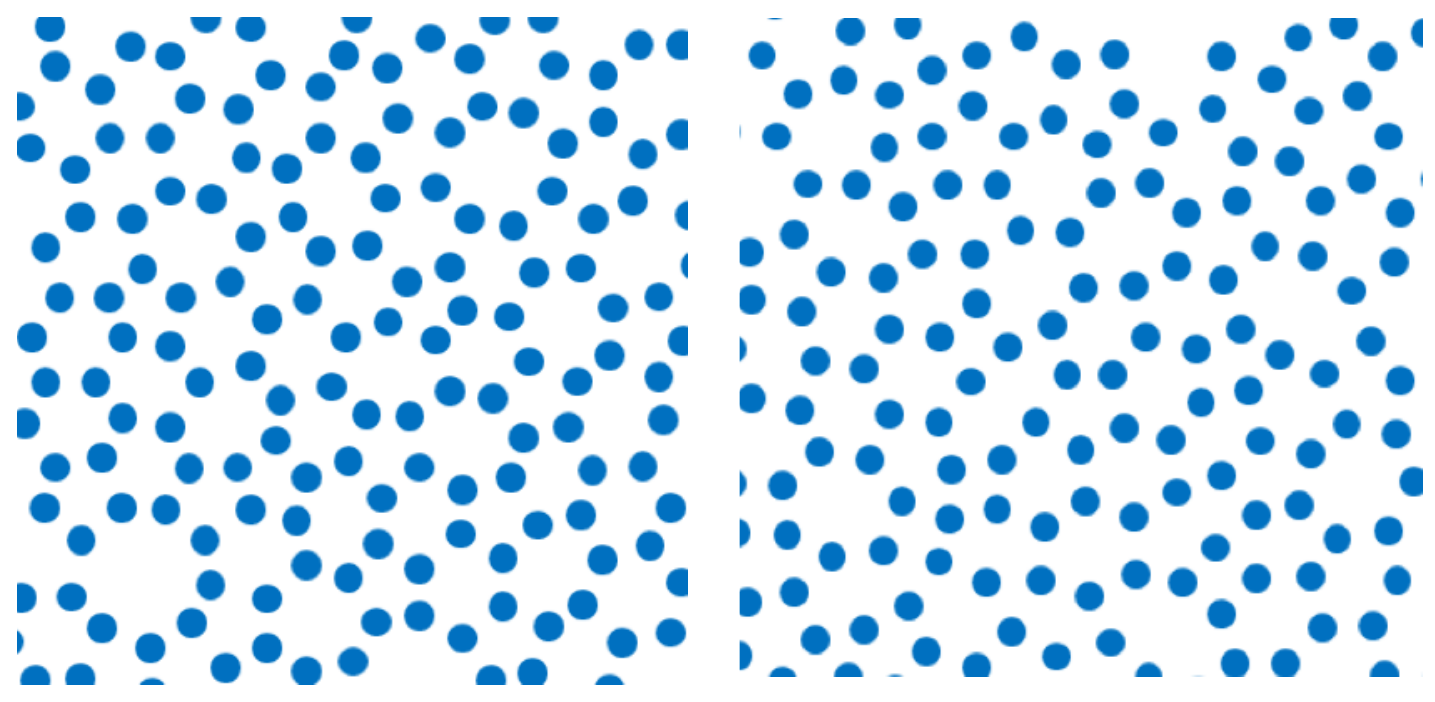}
\caption{Point configurations associated with the Si atoms derived based on the TEM images (left) and numerical network model (right). The two systems are visually very similar to one another.} \label{fig_4}
\end{figure*}

It can be seen again that these statistics agree well with one another. In particular, the $g_2$($r$) possesses an exclusion volume region, an average coordination number (the value of first peak) of 3 and quickly decays to unity, indicating no long-range order in the system. The P($r$) functions show narrow distributions, indicating small variations in the bond lengths. These results indicate that the local chemical order (i.e., Si-O bonds) induces strong local topological and geometrical constraints that are closely related to the observed disordered hyperuniformity in the system.

\section{High-Energy States in Amorphous Silica}
\begin{figure*}[ht]
\includegraphics[width=1\textwidth,keepaspectratio]{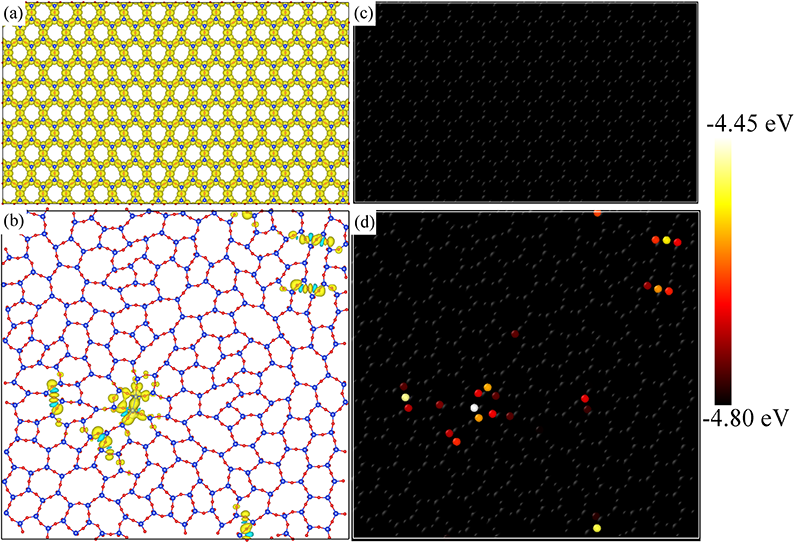}
\caption{Electron densities at the HOMO level of (a) the crystalline structure and (b) at the Fermi level of the DHU structure. The iso-surface value is 0.0023 $\times$10$^{-8}$ $e$/Bohr$^3$. The corresponding distributions of potential energies are shown in (c) and (d), respectively. The color bar shows a range of the distributions. A larger field of view for both systems are shown here than in the main text.} \label{fig_7}
\end{figure*}

\section{Stone-Wales Transformations Preserve Hyperuniformity}
In the main text, we provide an argument for the emergence of hyperuniformity in the amorphous 2D silica, i.e., the Stone-Wales transformations, which only induce local topological perturbations in the system, preserve hyperuniformity. In this section, we provide numerical evidence for this argument. 
\begin{figure*}[ht]
\includegraphics[width=1\textwidth,keepaspectratio]{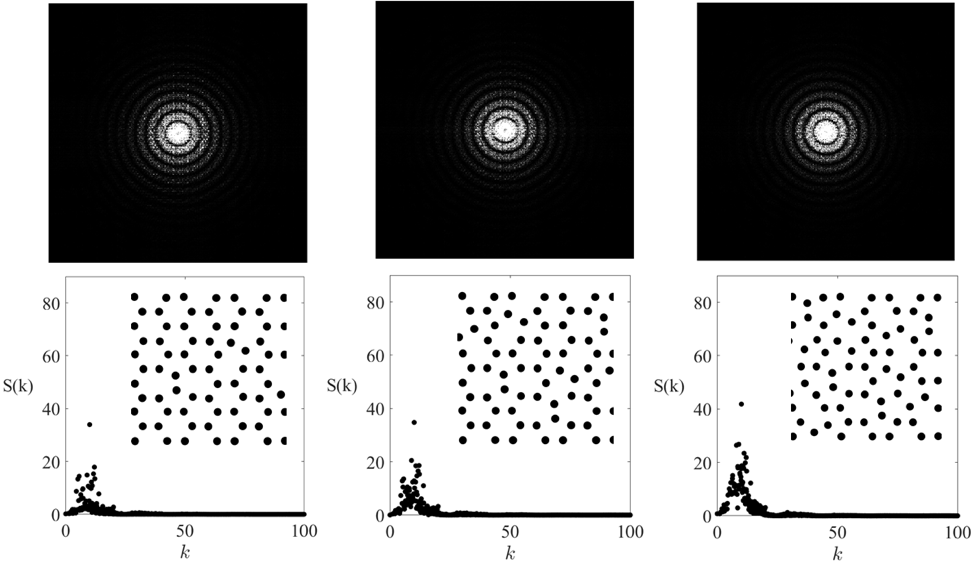}
\caption{Stone-Wales transformations preserve hyperuniformity. The figure shows the full spectral density (upper panels) and angular averaged spectral density (lower panels) for 2D distributions of Si atoms generated by random introducing different numbers of Stone-Wales defects. The concentration of the defects (see the text below for definition) is respectively 15\%, 45\%, and 70\% for the left, middle and right figures (see insets for configurations of Si atoms). It can be clearly seen that for all cases the systems remain hyperuniform, which indicates Stone-Wales transformations preserve hyperuniformity.} \label{fig_8}
\end{figure*}

In particular, we first generate a perfect honeycomb lattice packing of Si atoms (containing $\sim$ 1000 atoms), with the lattice vectors consistent with those defining the crystalline SiO2 structure. Next, we gradually introduce Stone-Wales defects in the system, by flipping randomly selected bonds, with constraints that a bond can be flipped only once. We define the concentration of the Stone-Wales defects as the number bonds affected by the Stone-Wales transformation over the total number of bonds in the system. The insets of Fig. S8 show typical configurations with different defect concentrations. 
Next, we compute the spectral density of the configurations containing different concentrations of Stone-Wales defects, which are shown in Fig. \ref{fig_8}. It can be clearly seen that for all concentrations we study (from 0\% to 75\%), the systems all remain hyperuniform, which indicates Stone-Wales transformations preserve hyperuniformity.
\bibliography{supplement}